\documentclass[sigconf,9pt]{acmart}

\usepackage{booktabs} 
\usepackage{multirow}
\usepackage{amsmath}
\usepackage{url}

\usepackage{color}
\usepackage{soul}
\usepackage{hyperref}

\usepackage[symbol]{footmisc}

 \hypersetup{
   colorlinks   = true, 
   urlcolor     = black, 
   linkcolor    = black, 
   citecolor   = black 
 }

\setcopyright{acmcopyright}

\renewcommand\footnotetextcopyrightpermission[1]{} 

\copyrightyear{2019} 
\acmYear{2019} 
\setcopyright{acmcopyright}
\acmConference[FPGA '19]{The 2019 ACM/SIGDA International Symposium on Field-Programmable Gate Arrays}{February 24--26, 2019}{Seaside, CA, USA}
\acmBooktitle{The 2019 ACM/SIGDA International Symposium on Field-Programmable Gate Arrays (FPGA '19), February 24--26, 2019, Seaside, CA, USA}
\acmPrice{15.00}
\acmDOI{10.1145/3289602.3293918}
\acmISBN{978-1-4503-6137-8/19/02}

\begin{document}
\title{Rapid Cycle-Accurate Simulator for High-Level Synthesis}


  \author{Yuze Chi, Young-kyu Choi,\footnotemark \hspace{0.05cm} Jason Cong, and Jie Wang}
  \affiliation{%
    \institution{Computer Science Department, University of California, Los Angeles}%
  }
  \email{{chiyuze, ykchoi, cong, jiewang}@cs.ucla.edu}%

\renewcommand{\shortauthors}{Yuze Chi, Young-kyu Choi, Jason Cong, and Jie Wang}

\begin{abstract}
A large semantic gap between the high-level synthesis (HLS) design and the low-level (on-board or RTL) simulation environment often creates a barrier for those who are not FPGA experts. Moreover, such low-level simulation takes a long time to complete. Software-based HLS simulators can help bridge this gap and accelerate the simulation process; however, we found that the current FPGA HLS commercial software simulators sometimes produce incorrect results. In order to solve this correctness issue while maintaining the high speed of a software-based simulator, this paper proposes a new HLS simulation flow named FLASH. The main idea behind the proposed flow is to extract the scheduling information from the HLS tool and automatically construct an equivalent cycle-accurate simulation model while preserving C semantics. Experimental results show that FLASH runs three orders of magnitude faster than the RTL simulation. 
\end{abstract}

%
%



\maketitle


\section{Introduction}

\footnotetext[1]{Corresponding author.}
\renewcommand{\thefootnote}{\arabic{footnote}}

Although FPGA has many promising features including power-efficiency and reconfigurability, the low-level programming environment makes it difficult for programmers to use the platform. In order to solve this problem, many high-level synthesis (HLS) tools such as Xilinx Vivado HLS \cite{Cong2011} and Intel OpenCL HLS \cite{Intel:SDK} have been released. These tools allow programmers to design FPGA applications with high-level languages such as C or OpenCL. This trend is reinforced by recent efforts on FPGA programming with languages of higher abstraction---such as Spark or Halide \cite{Segal2015, Sozzo2017, Yu2018}.

Even though such progress has been made on the design automation side, a large semantic gap still exists on the simulation side. Programmers often need to use low-level register-transfer level (RTL) simulators and try to map the result back to HLS. The result is often incomprehensible to those who are not FPGA experts. Moreover, such low-level simulation takes a very long time. Some work has been done to automate hardware probe insertion from the HLS source file \cite{Choi2017, Goeders2015, Monson2014, Verma2017}; however, this work requires regeneration of FPGA bitstream if there is a change in the debugging point, and the turnaround time is often in hours.

These problems can be partially solved by the software-based simulators provided by HLS tools. It takes little time to reconfigure the debugging points, and no semantic gap exists between the simulation and the design. However, a well-known shortcoming of these simulators is that most of them do not provide performance estimation. In addition, we found a critical deficiency---they sometimes provide \textit{incorrect} results. 

\begin{figure}[t]
\includegraphics[width=0.99\linewidth]{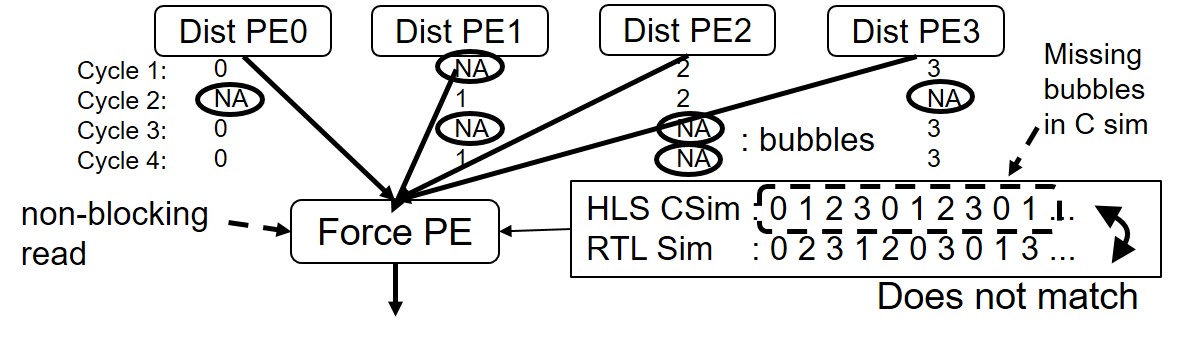}
\caption{Molecular dynamics simulation PEs \cite{Cong2016}}
\label{fig:namd}
\end{figure}

An example can be found in the molecular dynamics simulation \cite{Cong2016} (Fig.~\ref{fig:namd}). Multiple distance processing elements (Dist PEs) filter out faraway molecules above threshold and send them to Force PE. The pruned molecules will create a bubble (empty data) in the FIFO, and Force PE will process only the valid data (after non-blocking read) in the order they are received from any of the FIFOs. However, if the modules are instantiated in the order of (Dist PE1, PE2, ... Force PE) in the source file, Vivado HLS will finish the simulation of Dist PE1 first, followed by Dist PE2, and so on. As a result, by the time Force PE is simulated, the bubbles in the FIFOs are completely removed, and the Force PE output ordering can be entirely different from the actual result. If one was analyzing the DRAM access behavior from the HLS simulation output, the person would likely draw a wrong conclusion. 

Another problematic example can be found in the artificial deadlock situation \cite{Dai2014}, which occurs when the depth of the FIFO is smaller than the latency difference among modules (details in Section~\ref{sec:artificial-stall}). The first issue is that the HLS software simulator cannot detect the deadlock situation and proceeds as if there is no problem with the design. The second issue is that after we apply a transformation to remove the deadlock, the HLS tool cannot also simulate the amount of performance degradation (Section~{\ref{sec:cycle-accuracy}}) from the artificial stall (Section~{\ref{sec:artificial-stall}}). We also found a problem in the simulation of feedback loops where the feedback data is ignored by the HLS tool (Section~{\ref{sec:feedback}}).

\begin{table}[b]
\caption{Comparison of the software-based simulation of Xilinx Vivado HLS \cite{Xilinx:VivadoHLS} and Intel OpenCL HLS \cite{Intel:SDK}. Undesirable characteristics are in bold.}
\label{tab:xilinx_intel_comp}
\begin{tabular}{c|c c}
\hline
& Xilinx Viv HLS C Sim & Intel OpenCL HLS Sim \\
\hline
FIFO depth & \textbf{Unlimited} & Exact \\
Exec model & \textbf{Sequential} & Concurrent \\
Feedback & \textbf{Not supported} & Supported \\
Sim speed & $\sim$5 Mcycle/s & \textbf{$\sim$1 Mcycle/s} \\
Sim order & Deterministic & \textbf{Non-deterministic} \\
Max \# mods & No limit & \textbf{256} \\
Cycle-acc & \textbf{Not cycle-accurate} & \textbf{Not cycle-accurate} \\     
\hline
\end{tabular}
\end{table}

The primary reason for the incorrect simulation result is that HLS software simulators do not guarantee cycle accuracy. The comparison between the software simulator of the two most popular ({\cite{Lahti2018}}) commercial FPGA HLS tools, Xilinx Vivado HLS and Intel OpenCL HLS, is presented in Table~\ref{tab:xilinx_intel_comp}. Vivado HLS assumes unlimited FIFO depth which makes it difficult to accurately model FIFO fullness/emptiness. Also, their sequential simulation execution model prevents correctly simulating designs with feedback loops (Section~\ref{sec:feedback}). Intel OpenCL HLS simulates about 5X slower than Vivado HLS, but it correctly simulates the FIFO depth. The tool assigns a thread to each module for concurrent simulation; however, the execution order of the threads is not deterministic and may produce different results in different simulation runs for cases in Section~\ref{sec:problems}.

\begin{figure}[t]
\includegraphics[width=0.8\linewidth]{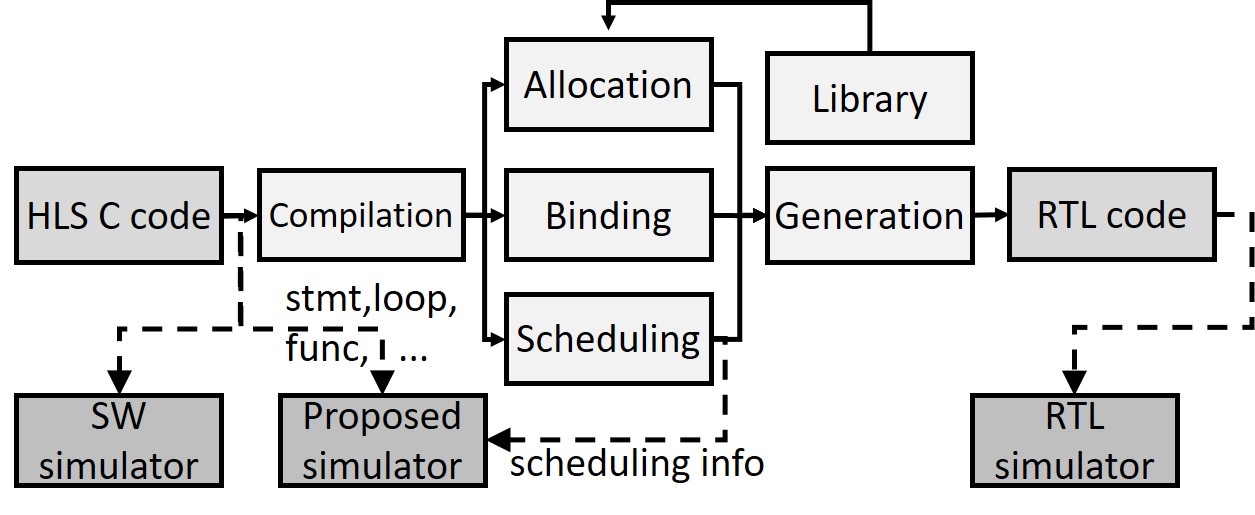}
\caption{HLS design steps \cite{Coussy2009} and simulation flows}
\label{fig:simflow}
\end{figure}

HLS design steps and conventional simulation flows are shown in Fig.~{\ref{fig:simflow}}. A software simulator runs fast but provides no cycle estimation and may have the correctness problem. An RTL simulator is accurate but runs slow since it incorporates low-level implementation details. Our solution to these problems is based on the idea that it may be possible to tackle both problems by simulating based on the scheduling information. It would be faster than the RTL simulation without the allocation / binding information and the component libraries; and it would solve the correctness problem of the software simulation and provide accurate performance estimation with its cycle-accuracy.

Although simulating solely based on the scheduler output (LLVM IR + scheduling information) is a possible option, we have instead decided to simulate in C syntax and augment it with scheduling information. The reason is that we wanted to raise the simulation abstraction level to further accelerate the simulation process and also make it easier for programmers to understand what is being simulated. To our knowledge, this is the first HLS-based simulation flow that takes such an approach.

By taking such an approach, however, several challenges were encountered (will be elaborated in Section~{\ref{sec:challenges}}). One problem is how to model high-level semantics such as functions and loops---as well as FIFO transactions and FIFO stalls---in a cycle-accurate fashion. Moreover, correctly simulating the task-level and pipelined parallelism that is inherent in hardware (and the corresponding RTL simulation) in sequential C semantics is a significant challenge.


In this paper we propose FLASH\footnotemark[1]---an HLS-based software simulation flow that addresses these challenges. We describe transformations that allow cycle-accurate simulation of communication and computation stages (will be explained in Section~{\ref{sec:challenges}}). Also, a method will be explained to simulate multiple levels of parallelism with C semantics. These steps will be described in Section~{\ref{sec:S2S_SIM}}.

\footnotetext[1]{FLASH: Fast, ParalleL, and Accurate Simulator for HLS}
\footnotetext[2]{Please cite [3], rather than this archive paper.}

We obtain the scheduling information from the HLS synthesis report and automatically generate a new simulation code based on the information. The new simulation code was made compatible with the conventional HLS software simulator for easy integration with the existing tool. The overall flow is described in Section~{\ref{sec:overall}}.


Our current initial version is based on Vivado HLS, but we hope to extend our work to Intel HLS if the tool provides detailed internal scheduling information in the future.

This paper is an extended version of \cite{Chi2019}, which has been accepted for publication in FPGA'19.\footnotemark[2]

\section{Related Work}

Work in \cite{Choi2017, Goeders2015, Monson2014, Verma2017} describe frameworks that allow users to specify debugging points in high-level language and synthesize hardware probes into the FPGA for analysis. They can be categorized into work that has more focus on verifying functional correctness {\cite{Monson2014,Goeders2015}} and work that has more focus on extracting performance-related parameters \cite{Choi2017, Verma2017}. 
Work in {\cite{Goeders2015}} describes how to record and replay the execution of optimized HLS-generated circuits.
Work in {\cite{Monson2014}} explains how to combine multiple signals to reduce trace buffer size. 
HLScope \cite{Choi2017} describes an in-FPGA monitoring flow that extracts cycle information from FPGA designs written in C.
Work in {\cite{Verma2017}} is based on OpenCL and measures stall latency and monitors memory access patterns by utilizing trace buffers to store an event's timestamp.
However, these hardware-based debuggers typically requires hours of initial overhead for bitstream generation.

There are several SystemC simulators \cite{Chung2014, Schmidt2017} that can achieve cycle-accuracy for the source code that has explicit scheduling information specified by the programmer, but this may be too difficult for non-experts. Our flow, on the other hand, achieves cycle-accuracy for a HLS C source code that does not have such user-defined scheduling information.

There are also other HLS-based software simulators.
The LegUp HLS \cite{Canis2013} simulator provides speedup prediction based on the profiling of the source code and the execution cycle from its synthesis result. HLScope+ \cite{Choi2017+} describes a method to extract cycle information that is hidden by HLS abstraction and uses Vivado HLS C simulation to predict the performance for applications with dynamic behavior. These works, however, do not guarantee cycle-accuracy.

\section{Problem Description and Motivating Examples}
\label{sec:problems}

In this section we describe three classes of problems that cause current HLS tools to produce incorrect software simulation result. The problems are demonstrated with motivating examples in the literature.

\subsection{Incorrect~Data~Ordering~with~Multiple~Paths}
\label{sec:dataorder}

Suppose a PE is reading data in a non-blocking fashion from multiple PEs through FIFOs as in the molecular dynamics simulation example (Fig.~\ref{fig:namd}~\cite{Cong2016}) in the introduction. If a bubble exists in a FIFO, the data consumer PE will skip the FIFO and proceed to read from the next FIFO. In software simulation, however, if the data producer PEs are instantiated in the source file before the consumer PE, Vivado HLS will simulate the data producer PEs completely before moving on the next one. This effectively removes all bubbles in the FIFO, and the order of output from the data consumers in the software simulation result will be different from the actual execution. In the Intel HLS, the simulation order of the data producers is undetermined, and thus there is no guarantee that the bubbles in the simulated result will exactly match the actual execution.

\subsection{Artificial Deadlock and Stall}
\label{sec:artificial-stall}



\begin{figure}[t]
\includegraphics[width=0.99\linewidth]{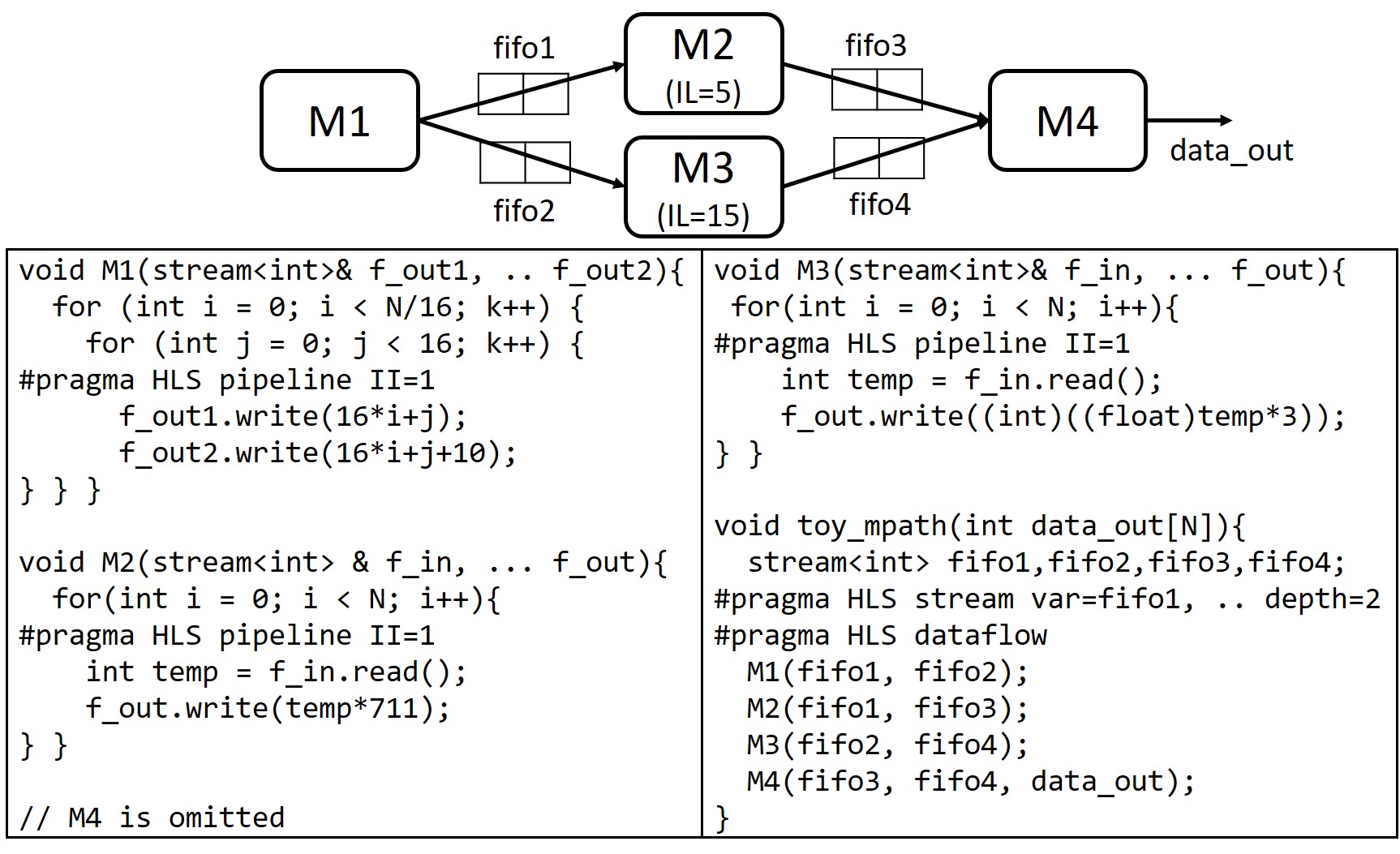}
\caption{Structure and code for motivating example \texttt{toy\_mpath}}
\label{fig:toy_mpath}
\end{figure}

\begin{figure}[t]
\includegraphics[width=0.95\linewidth]{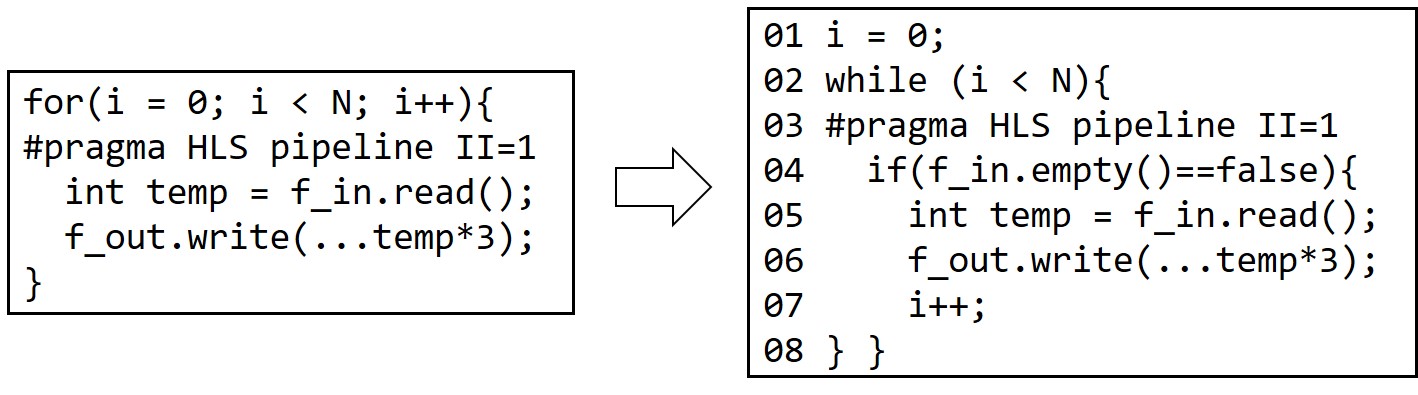}
\caption{Source-to-source code transformation to avoid artificial deadlock for \texttt{M3} in Fig.~\ref{fig:toy_mpath}}
\label{fig:ad_transf}
\end{figure}

Consider an example in Figure~\ref{fig:toy_mpath} where the module \texttt{M2} has a latency of 5 and \texttt{M3} has a latency of 15. All FIFOs have a depth of 2. 
After \texttt{M2} has produced two output elements, \texttt{M4} cannot consume any of them because \texttt{fifo4} is still empty due to the long latency of \texttt{M3}. 
Due to the back-pressure from \texttt{M2} and \texttt{fifo3}, \texttt{fifo1} becomes full. 
Then \texttt{M1} will stop producing output to \texttt{fifo2} because \texttt{fifo1} and \texttt{fifo2} have to be written in the same cycle.
\texttt{fifo2} will eventually become empty, which blocks the pipeline of \texttt{M3}.
Then none of the modules can do any further useful work, and the circuit deadlocks.
This is called an artificial deadlock \cite{Dai2014}.
The deadlock is caused by the mismatching latency of multiple paths and the small FIFO depth. 
This can be observed in real applications, such as the dataflow-based architecture for stencil computations in \cite{Chi2018} that contains various modules and FIFOs with different latencies and depths.


The problem is that software-based HLS simulators ignores the latency of a module. It will simulate each iteration of a loop as if the data is instantaneously passed from input to output. Thus Vivado HLS will proceed with the simulation as if the deadlock has not happened. Intel HLS compiler avoids the deadlock problem by automatically increasing the FIFO depth; however, this creates a new problem of mismatch between what is simulated and synthesized.



The second problem was found after we applied code transformation to avoid the deadlock. Figure~\ref{fig:ad_transf} shows the transformation for \texttt{M3} in Figure~\ref{fig:toy_mpath}. If the input FIFO is empty, a bubble is inserted into the pipeline (line~4)---this allows the pipeline to keep processing the already-read data even if there is no additional input. The deadlock situation is removed since \texttt{M4} can now receive the output from \texttt{M3}.

Even though the deadlock was avoided, however, the modules still have to wait for the data to be flushed. This causes a delay that we call \textit{artificial stall}. Since HLS tools do not consider the delay due to the latency of a module, such performance degradation cannot be simulated.

\subsection{Missing Data from Feedback Path}
\label{sec:feedback}

\begin{figure}[t]
\includegraphics[width=0.82\linewidth]{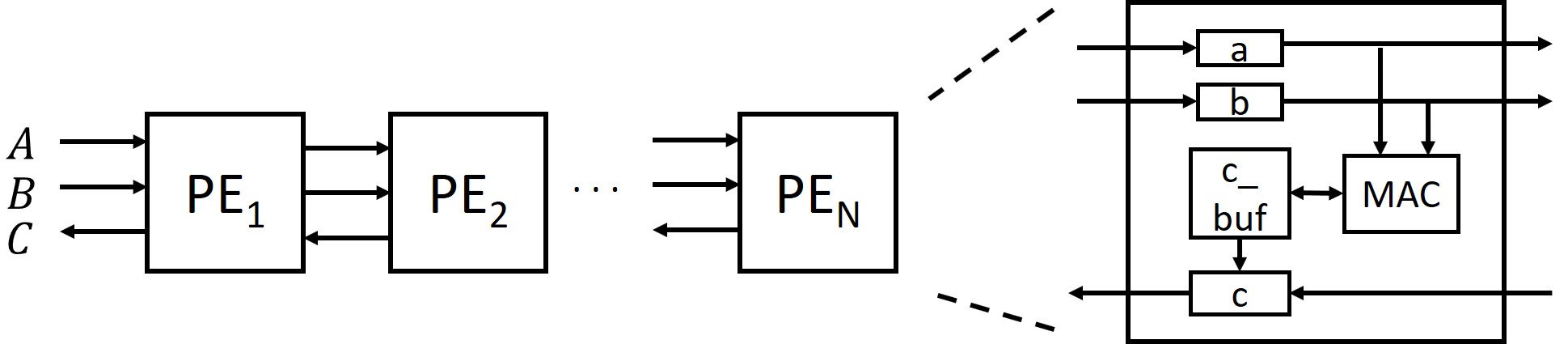}
\caption{Matrix multiplication with linear systolic array architecture}
\label{fig:sysmmul}
\end{figure}

As mentioned previously, Vivado HLS simulates the functions in the order they are instantiated in the source code. This causes a problem if a feedback path exists that passes data from later instantiated functions to earlier ones. At the time earlier functions are simulated, the data would not be available. As a result, Vivado HLS simulates the program as if the feedback FIFOs are always empty. Intel HLS can simulate the feedback data from blocking read correctly, because a thread simulating each module can wait for others to pass the data---although it is not guaranteed that the feedback data from non-blocking read will arrive at the right timing.

We demonstrate this problem with matrix multiplication example ($C=A\times B$) in linear systolic array architecture \cite{Cong2018,Jang2005}. As shown in Fig.~\ref{fig:sysmmul}, each PE computes one column of the matrix $C$ ($C_{ij}\mathrel{{+}{=}}A_{ik}*B_{kj}$). Data from the matrix $A$ and $B$ are fed into the array in the forward direction, while the results of matrix $C$ are collected in the backward direction. If the modules are instantiated in the order of $PE_{1}$, $PE_{2}$, ..., and $PE_{N}$, Vivado HLS will simulate $PE_{1}$ assuming the FIFO for $C$ is always empty, and this will cause the tool to produce incorrect results.

\section{Problem Statement and Challenges}
\label{sec:challenges}

The data ordering problem (Section~{\ref{sec:dataorder}}) can be solved if the simulator models the FIFO data transaction (read/write) and the FIFO stall (empty/full) in a cycle-accurate fashion. The artificial deadlock problem (Section~{\ref{sec:artificial-stall}}) requires modules to initiate FIFO read and write at the timing that reflects the computation latency. In other words, it requires cycle-accurate modeling of \textit{computation stages}, which we define as the computation latency between pairs of FIFO read and FIFO write. The feedback problem (Section~{\ref{sec:feedback}}) does not occur if the FIFO read in the feedback path is simulated after the FIFO write. 

Thus, the problem is stated as follows: given a source code and its scheduling information, we need a simulator that models the communication and the computation stages in a cycle-accurate manner. The simulator also must produce correct output data.


In addition to this main requirement, the simulator should be able to provide the execution cycles of each module to help programmers apply performance optimization. Also, if the modules deadlock, the simulator should provide the content of the internal registers for debugging purpose. Moreover, the simulation code should be semantically similar to the source code as much as possible (as opposed to being a low-level code such as RTL), so that users can easily understand what is being simulated.

With such complicated requirements, several challenges arise:
\begin{itemize}
    \item \textbf{Challenge 1 : Cycle-accurate simulation} \\
It is difficult to discover the exact cycle when statements are executed since the information given by the HLS tool is very limited. Intel OpenCL HLS only provides loop initiation interval (II). Vivado HLS provides slightly more information---such as the module's finite-state machine (FSM) state when FIFO read or write is performed. However, for computation statements, it is difficult to find the exact cycle, because Vivado HLS only provides lists of LLVM IR and the corresponding FSM states. Mapping such low-level representation back to the original C code is a difficult task.\\
Also, even if the schedule of all operations are known, the simulator has to \textit{selectively} execute statements that correspond to a particular FSM state at each cycle. Moreover, the content of the variables in the previous state has to be available, and the updated variables have to be stored for the next state simulation. 
	\item \textbf{Challenge 2 : Simulation of parallelism} \\
RTL is an inherently parallel language---it has multiple levels of parallelism including task-level parallelism and pipelined parallelism. On the other hand, pure C is written in a sequential form. The challenge is in transforming C into a form that can simulate the concurrency.
	\item \textbf{Challenge 3 : FIFO communication and pipeline stall} \\
In RTL simulation, a full or empty signal from FIFO can halt an FSM. An equivalent software simulator would also need to mimic this behavior based on the status of the FIFOs. Also, a deadlock would need to be detected if all pipelines can no longer make any progress.
	\item \textbf{Challenge 4 : Loop and function simulation} \\
We would need to construct an equivalent model of high-level semantics, such as loops and functions.
\end{itemize}

\section{Automated Code Generation for Rapid Cycle-Accurate Simulation}
\label{sec:S2S_SIM}

In this section, we provide a solution to each challenge in Section~\ref{sec:challenges} and describe our proposed automated simulation code generation flow. For illustration, we will use the \texttt{toy\_mpath} example (Fig.~{\ref{fig:toy_mpath}}) after applying the deadlock avoidance transformation discussed in Section~{\ref{sec:artificial-stall}}.

\subsection{Cycle-Accurate Simulation}

\begin{figure}[t]
\includegraphics[width=0.99\linewidth]{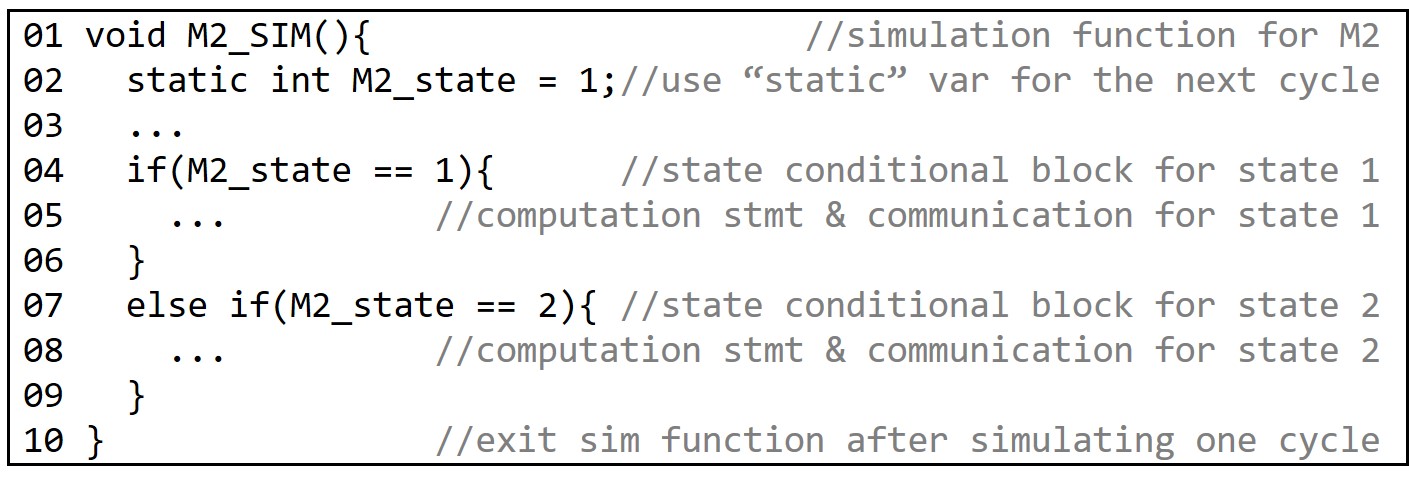}
\caption{Simulation function structure for cycle-accurate simulation}
\label{fig:simfunc}
\end{figure}

For cycle-accurate simulation, we declare an FSM state variable for each module and copy statements to the conditional block that correspond to its simulated state. An example can be found for \texttt{M2} module in lines 4--9 of Fig.~\ref{fig:simfunc}. Only the statements for a single cycle are simulated and then the simulation function exits. The contents of the variables are restored and saved regardless of simulation function entrance or exit by using \texttt{static} variables (line 2).

Regardless of the exact cycle a computation statement is simulated, we exploit the fact that the behavior observed from outside the module (including the module's computation stage) would be the same as long as the inter-module FIFO communication is simulated at the correct cycle. Thus, even if the schedule of a module's computation statement is unknown, we can assign an arbitrary state that does not violate the timing causality with the cycle-known FIFO communication that has dependency with the computation statement.
We assign states to the computation statements based on as-soon-as-possible scheduling policy to reduce the number of pipelined shift registers (Section~\ref{sec:pipe_model}). The simulation of computation statements and FIFO communication will be further explained in Section~\ref{sec:pipe_model} and Section~\ref{sec:fifo_model}, respectively.

\subsection{Simulation of Parallelism}

\subsubsection{Pipelined Parallelism}
\label{sec:pipe_model}

\begin{figure}[t]
\includegraphics[width=0.99\linewidth]{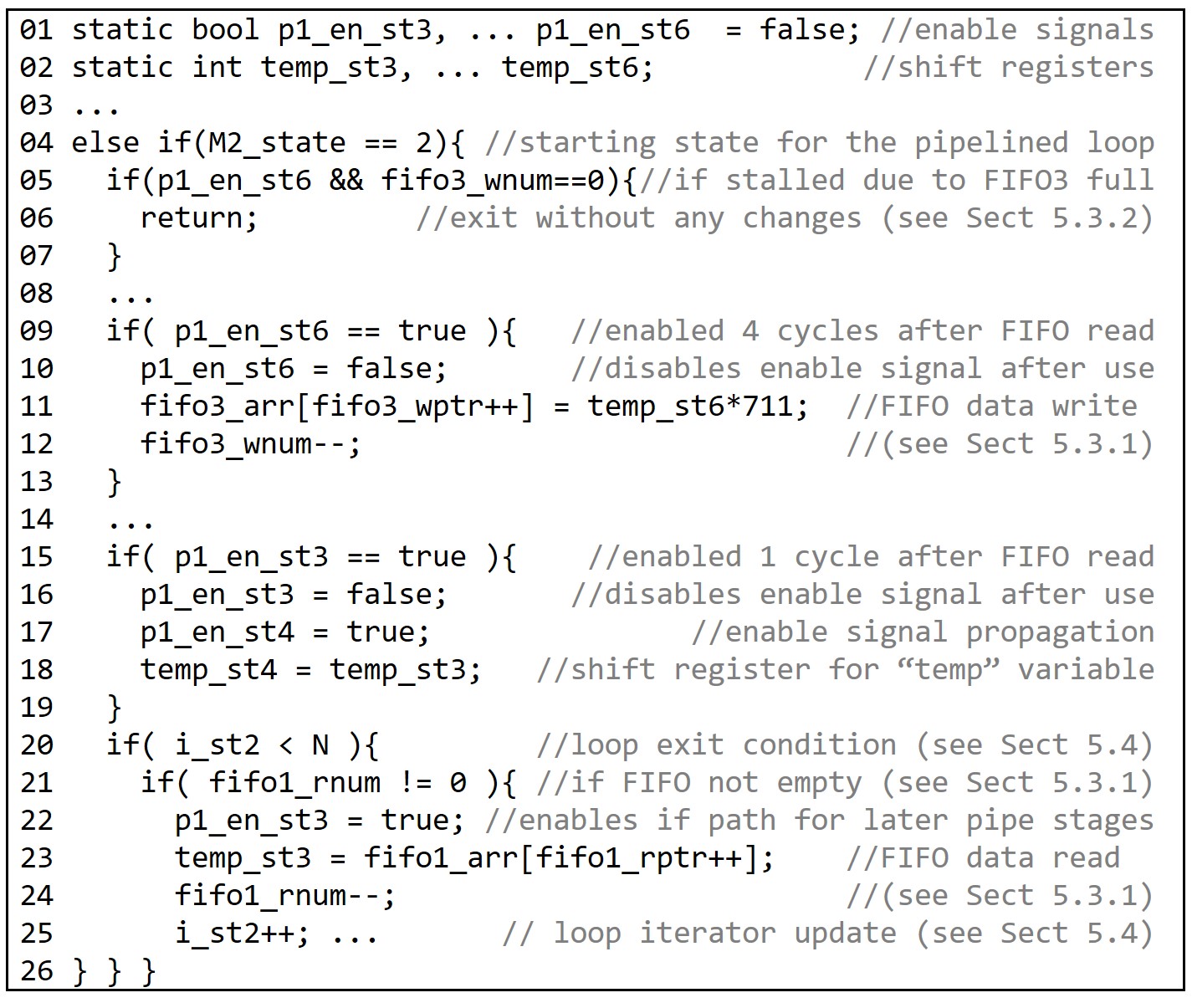}
\caption{Code transformation to model cycle-accurate, pipelined parallelism (\texttt{M2} in Fig.~\ref{fig:toy_mpath})}
\label{fig:pipeline_code}
\end{figure}

In a pipelined loop, different iterations are executed in parallel in a single FSM state. The parallel factor is same as the loop iteration latency (IL, also called pipeline depth). To simulate such parallelism, we need to keep multiple copies of the same variable for each pipelined stage. For example, the "temp" variable in \texttt{M2} (Fig.~\ref{fig:toy_mpath}) is copied through the pipeline like shift registers (line 17 of Fig.~\ref{fig:pipeline_code}). We perform liveness analysis on each pipelined variable to reduce its number. Next, instead of placing the computation for each pipeline stage in a corresponding M2\_state conditional block as in Fig.~\ref{fig:simfunc}, we place all computation in a single M2\_state conditional block as shown in lines 4--26 of Fig.~\ref{fig:pipeline_code}. This transformation allows us to effectively simulate the pipelined parallelism. If II is larger than 1, the computation at state $i$ is placed at the state conditional block of $i\%II$.

It is important to note that the order of each pipeline stage has been \textit{reversed} (st6, ... st3, st2). This limits the content of shift register to be copied to the immediate next state only in a single cycle. Also, in order to invalidate a pipeline bubble (from the artificial deadlock avoidance transformation in Section~\ref{sec:artificial-stall}), we propagate the enable signal through the pipeline stages (line 17 and 22).

\subsubsection{Task-Level Parallelism}
\label{sec:taskpar}

\begin{figure}[t]
\includegraphics[width=0.9\linewidth]{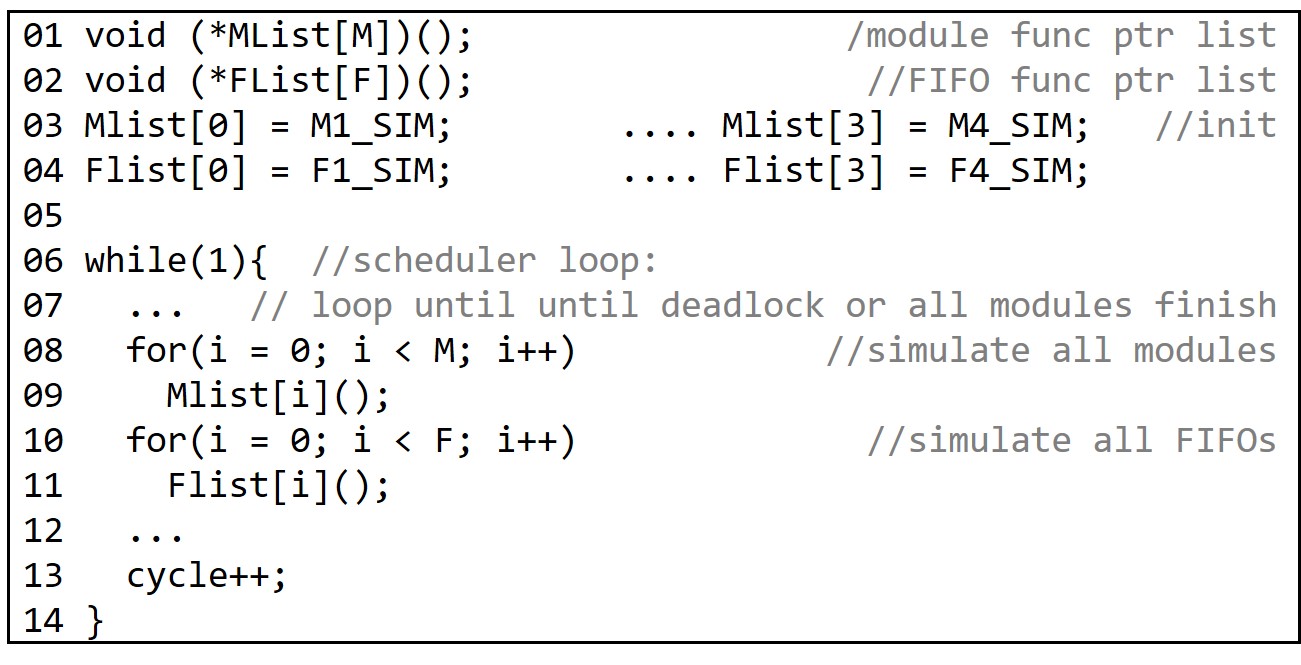}
\caption{Module/FIFO simulation scheduler to model task-level parallelism}
\label{fig:taskpar_code}
\end{figure}

The task-level parallelism is simulated by processing one cycle of all modules and FIFOs in a round-robin fashion. This is processed in the scheduler loop in line~6-14 of Fig.~\ref{fig:taskpar_code}. It is composed of module (line~8-9) and FIFO (line~10-11) simulation loop.

It is possible that different order of the module and FIFO simulation loop leads to different output---for example, depending on if the data producer PE is simulated before or after the consumer PE. A way to avoid this problem will be discussed in Section~{\ref{sec:fifo_comm}}.


\subsection{FIFO Simulation}
\label{sec:fifo_model}

\subsubsection{FIFO Communication}
\label{sec:fifo_comm}

The FIFO is implemented as a circular buffer with read/write pointers (\texttt{fifo\_rptr} and \texttt{fifo\_wptr}) and an array (\texttt{fifo\_arr}). The array length is set to FIFO buffer size (\texttt{FIFO\_SIZE}) plus one, because one buffer space is kept empty in circular buffer implementation \cite{Cormen2005}. Also, we declare \texttt{fifo\_rnum} and \texttt{fifo\_wnum} variables to denote the number of data and buffer space available in the FIFO. FIFO reads and writes in the source code are transformed based on Table~{\ref{tab:fifo_code}}. For example, the FIFO write in \texttt{M2} (fifth line of \texttt{M2} in Fig.~{\ref{fig:toy_mpath}}) would be transformed to: ($fifo3\_arr$ $[fifo3\_wptr\mathrel{{+}{+}}]$ $= temp\_st6*711;$ $fifo3\_wnum\mathrel{{-}{-}};$) (line~11-12 of Fig.~{\ref{fig:pipeline_code}}). 

\begin{figure}[t]
\includegraphics[width=0.99\linewidth]{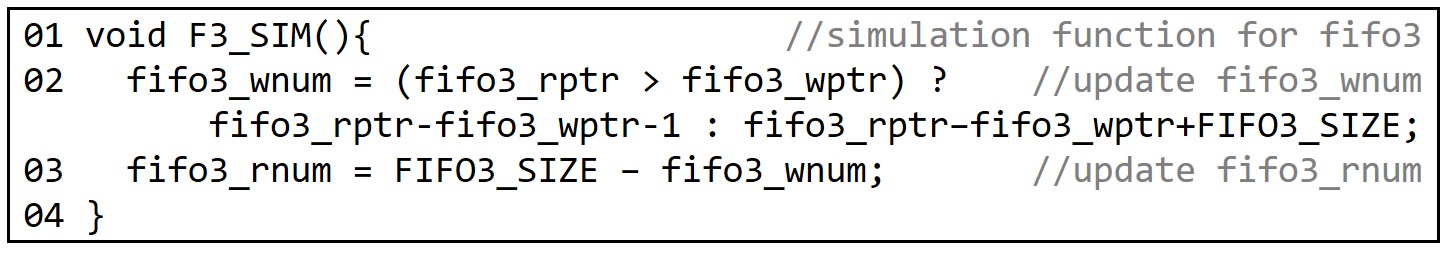}
\caption{Simulation code for \texttt{fifo3}}
\label{fig:fifo_code}
\end{figure}

In addition to decreasing the number of buffer space ($fifo3$ $\_wnum\mathrel{{-}{-}};$) for FIFO write, we would need to increase the number of available data ($fifo3\_rnum\mathrel{{+}{+}};$). However, this process is delayed until the FIFO simulation loop (line~10-11 of Fig.~{\ref{fig:taskpar_code}}, and more details in Fig.~\ref{fig:fifo_code}). The reason is to ensure that simulating data producer PE earlier than the consumer PE (in the module simulation loop in line~8-9 of Fig.~{\ref{fig:taskpar_code}}) does not allow transfer of data through the FIFO in the same cycle (1 cycle latency is needed).

\begin{table}[h]
\caption{Code transformation for FIFO communication}
\label{tab:fifo_code}
\begin{tabular}{l|l}
\hline
HLS source code & Transformed simulation code \\
\hline
fifo.empty() & fifo\_rnum == 0\\
fifo.full() & fifo\_wnum == 0\\
data = fifo.read() & \small data = fifo\_arr[fifo\_rptr$\mathrel{{+}{+}}$]; fifo\_rnum$\mathrel{{-}{-}}$;\normalsize\\
fifo.write(data) & \small fifo\_arr[fifo\_wptr$\mathrel{{+}{+}}$] = data; fifo\_wnum$\mathrel{{-}{-}}$;\normalsize\\
\hline
\end{tabular}
\end{table}

\subsubsection{Pipeline Stall Modeling}

If a pipeline stall condition is met, none of the statements should be simulated at the current state. Thus, the stall condition should be placed at the beginning of a state conditional block. This will make the simulation function to exit without changing any variables. After applying the artificial deadlock avoidance transformation, FIFO read no longer causes the stall, but FIFO write will. The stall condition is met when the FIFO is full and when the state for the FIFO write statement has been enabled. For example, the pipeline stall condition that corresponds to FIFO write in line~11 of Fig.~\ref{fig:pipeline_code} would be : $if$($p1\_en\_st6$ $\&\&$ $fifo3\_wnum==0$). This condition has been added to line~5-7 of Fig.~{\ref{fig:pipeline_code}}.

Note that our tool can detect a deadlock by checking if no state transition occurs (stalled) in any modules and no data transaction occurs in any FIFOs. This may happen if the user decides not to incorporate the artificial deadlock avoidance method (Section~\ref{sec:artificial-stall}).

\subsection{Loop and Function Simulation}

\begin{figure}[t]
\includegraphics[width=0.6\linewidth]{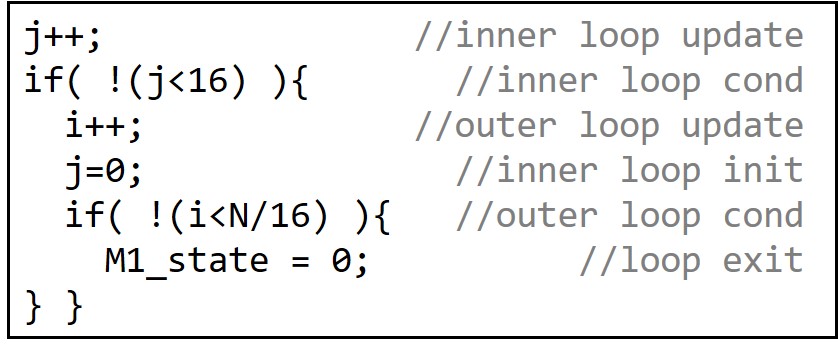}
\caption{Loop condition and update for flattened loop in \texttt{M1} of Fig.~\ref{fig:toy_mpath}}
\label{fig:loop_proc}
\end{figure}

Simulation of statements inside a pipelined loop has been discussed in Section~{\ref{sec:pipe_model}}. For the loop initialization statement, it is simulated upon entering the first state of a loop. The loop update expression is simulated at each iteration of a loop. If the loop condition is met after the update, state transition for loop exit occurs. For a flattened loop (e.g., \texttt{M1} in Fig.~\ref{fig:toy_mpath}), the update and the loop condition check is performed starting from the innermost nested loop, as illustrated in Fig.~\ref{fig:loop_proc}.

A function call is simulated by sending a module enable signal to the scheduler loop (Fig.~{\ref{fig:taskpar_code}}). Next, the function argument values are copied into the newly called module.

\section{Overall Flow}
\label{sec:overall}

\begin{figure}[t]
\includegraphics[width=0.99\linewidth]{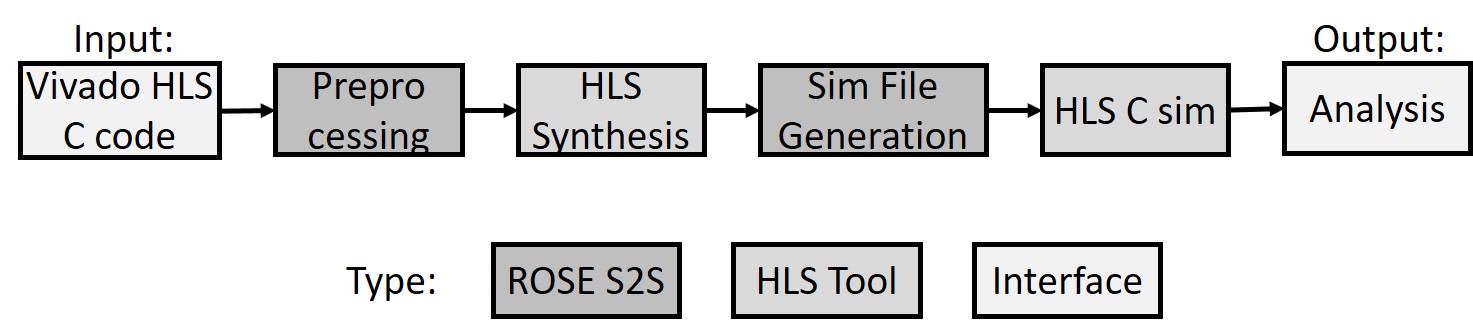}
\caption{Overall simulation framework of FLASH}
\label{fig:overall}
\end{figure}

The overall simulation framework of FLASH is shown in Fig.~\ref{fig:overall}. Given an input Vivado HLS C code, we apply an optional preprocessing step of transforming pipelined loops to avoid artificial deadlock (Section~\ref{sec:artificial-stall}). Also, some labels are added to easily identify loops and functions. The transformation step uses the APIs in the ROSE compiler infrastructure {\cite{Rose}}. The transformed code is fed into the Vivado HLS for synthesis. Based on the scheduling report given by the HLS tool, the input code is automatically transformed for rapid software simulation (Section~\ref{sec:S2S_SIM}). The simulation code has been made compatible with the Vivado HLS software simulator for easy integration with the existing tool. As a final output, our flow currently provides the number of cycles consumed in each module. As a future effort it will be enhanced to provide both functional debugging support (e.g., data dump, triggers), and performance debugging support (e.g., module stall analysis).

\section{Experimental Results}
\label{sec:exp}

\subsection{Experimental Setup}

For HLS tool, we use Vivado HLS 2018.2 {\cite{Xilinx:VivadoHLS}}.
For platform, we target the ADM-PCIE-KU3 board {\cite{Alphadata:KU3}} with Xilinx's Ultrascale KU060 FPGA \cite{Xilinx:UltraScale}. The target clock frequency is 250MHz.
The simulation is conducted with a server node that has Intel Xeon Processor E5-2680 {\cite{Intel:XeonE5}} and 64GB of DRAM. The simulation files were compiled with --O3 flag.

The experiment is performed on \texttt{toy\_mpath} (Fig.~\ref{fig:toy_mpath}) and three dataflow benchmarks: stencil \cite{Chi2018}, molecular dynamics simulation \cite{Cong2016} (Fig.~\ref{fig:namd}), and matrix multiplication \cite{Cong2018} (Fig.~\ref{fig:sysmmul}).

\subsection{Execution Time}

As mentioned in Section~\ref{sec:overall}, preprocessing, HLS synthesis, and simulation file generation steps are needed to prepare the files for the proposed simulation. The time breakdown of the steps is presented in Table~\ref{tab:s2s_time}.

\begin{table}[h]
\caption{Simulation preparation time breakdown (preprocessing, HLS synthesis, and simulation file generation: Fig.~\ref{fig:overall})}
\label{tab:s2s_time}
\begin{tabular}{c|c c c|c}
\hline
Benchmark & Preproc & HLS Synth & SimFile Gen & Total \\
\hline
\texttt{Toy\_mpath} & 7.1s & 24s & 7.5s & 39s \\
\texttt{Stencil} & 15s & 60s & 22s & 97s \\  
\texttt{MD\_sim} & 8.0s & 35s & 11s & 54s \\
\texttt{Mat\_mul} & 8.1s & 31s & 10s & 49s \\
\hline
\end{tabular}
\end{table}

The simulation time comparison among Vivado HLS C simulation, Vivado HLS RTL simulation, Intel OpenCL HLS simulation (using Quartus 18.0 \cite{Intel:Quartus}), and our FLASH simulation flow is presented in Table~\ref{tab:sim_time}. FLASH is about 1,390X (=1,570/1.13) faster than the RTL simulation. This confirms our initial speculation that simulating based on the scheduling information will result in much faster speed, since the simulation is not slowed by the resource allocation / binding information or the component library that exist in RTL simulation.

\begin{table}[h]
\caption{Simulation time comparison among Vivado HLS C simulation, Vivado HLS RTL simulation, Intel OpenCL HLS simulation, and FLASH simulation}
\label{tab:sim_time}
\begin{tabular}{c|c c c c}
\hline
Benchmark & V C Sim & V RTL Sim & I OCL Sim & FLASH\\
\hline
\multirow{2}{*}{\texttt{Toy\_mpath}} & 0.602s & 492s & 4.60s & 0.570s \\
& (1.00X) & (817X) & (7.64X) & (0.947X) \\
\multirow{2}{*}{\texttt{Stencil}} & 1.46s & 113s & 2.63s & 1.25s \\  
 & (1.00X) & (77.4X) & (1.80X) & (0.856X) \\
\multirow{2}{*}{\texttt{MD\_sim}} & 0.0547s & 100s & 0.0921s & 0.0677s \\
 & (1.00X) & (1,830X) & (1.68X) & (1.24X) \\
\multirow{2}{*}{\texttt{Mat\_mul}} & 0.0539s & 192s & 0.201s & 0.0810s \\
& (1.00X) & (3,560X) & (3.73X) & (1.50X) \\
\hline
AVG & (1.00X) & (1,570X) & (3.71X) & (1.13X) \\
\hline
\end{tabular}
\end{table}

Since our flow reflects the scheduling information, we can expect some slowdown compared to the Vivado HLS C simulation. This is noticeable in \texttt{Mat\_mul}, where the frequent FIFO stall (Table~{\ref{tab:accuracy}}) lengthens the simulation process. \texttt{MD\_sim} has a long simulation time due to the deep pipeline (55)---the overhead of copying shift registers and enable signals (Section~{\ref{sec:pipe_model}}) for pipeline stages becomes relatively large. However, it is interesting to note that for \texttt{Toy\_mpath} and \texttt{Stencil}, FLASH was even faster than the Vivado HLS C simulation. This suggests that there was an unexpected factor which has negated the simulation speed overhead of the proposed flow. We found that this is largely attributed to the fact that Vivado HLS can allocate unlimited FIFO buffer for C simulation (Table~{\ref{tab:xilinx_intel_comp}}). To model FIFO, the Vivado HLS C simulator uses the C++ Standard Template Library (queue.h), which incurs the overhead of dynamically allocating buffer and copying its content. For example, the C simulation time of \texttt{Toy\_mpath} reduces from 0.602s to 0.076s if we replace FIFO library calls with fixed-size arrays (array size is set to the number of total FIFO elements written). FLASH simulation flow does not have this problem, because the FIFO library calls have been replaced with array-based communication (Section~{\ref{sec:fifo_model}}). The average slowdown of FLASH compared to the Vivado HLS C simulation is 1.13X.

Please note that in our initial research stage, we also evaluated a similar flow with SystemC. However, the overhead in SystemC simulation environment was causing a 2-3X slowdown compared to the proposed C-based flow, which motivated us to follow the current approach.

\subsection{Accuracy}
\label{sec:cycle-accuracy}

As explained in Section~{\ref{sec:challenges}}, the correctness problem can be solved by simulating in a cycle-accurate manner.
The data value and the data ordering has been verified by comparing the output of FLASH simulator with that of the RTL simulator.

In Table~{\ref{tab:accuracy}}, we compare the cycle estimation accuracy with Vivado HLS synthesis report after we specify the maximum loop bound for each loop. We were not able to provide comparison with Intel HLS since the tool does not provide cycle estimate. The estimation error rate is small for \texttt{Stencil}, because \cite{Chi2018} has built-in mechanism to allocate adequate buffers. For the rest of the benchmarks, we have applied a small (1--2) FIFO depth (an example was shown in Fig.~{\ref{fig:toy_mpath}}). This causes FIFO buffer to be frequently full and empty and leads to worse performance than what HLS tool has predicted. Our flow, on the other hand, simulates in a cycle-accurate fashion and accurately estimates such performance degradation.

\begin{table}[h]
\caption{Total execution cycle predicted by Vivado HLS synthesis report and FLASH, and its error rate compared to the RTL-simulated result}
\label{tab:accuracy}
\begin{tabular}{c|c c c}
\hline
Benchmark & RTL sim & Vivado HLS & FLASH\\
\hline
\multirow{2}{*}{\texttt{Toy\_mpath}} & 4,500,010 & 2,000,016 & 4,500,010 \\
& - & (-56\%) & (0\%) \\
\multirow{2}{*}{\texttt{Stencil}} & 524,309 & 524,299 & 524,309 \\
& - & (\textasciitilde0\%) & (0\%) \\
\multirow{2}{*}{\texttt{MD\_sim}} & 12,089 & 10,498 & 12,089 \\
& - & (-13\%) & (0\%) \\
\multirow{2}{*}{\texttt{Mat\_mul}} & 330,006 & 131,075 & 330,006\\
& - & (-60\%) & (0\%) \\
\hline
AVG & - & (-32\%) & (0\%) \\
\hline
\end{tabular}
\end{table}

\section{Concluding Remarks}

By simulating based on the scheduling information, we were able to solve the correctness issue of the software simulators and also provide accurate performance estimation. Also, simulating without allocation / binding information and component libraries allowed us to achieve three orders of magnitude faster speed compared to the RTL simulators. We have described an automated code generation flow that enables this new simulation flow.

We hope that the promising result presented in this work will motivate HLS commercial tool industry to provide additional routine that simulates based on the scheduling information only. This will substantially decrease the validation time of the customers who wish to rapidly estimate cycle-accurate performance, obtain correct output data, or detect possible deadlock situations.

As a future work, we will continue to widen the range of benchmarks so that the transformation flow will be robust enough to accommodate any Vivado HLS input code. We hope to include the Intel HLS flow if their tool's synthesis report provides detailed schedule information in the future. Also, we will enhance the output analysis stage to provide better functional and performance debugging support. In addition, we plan to add parallelization using Pthread/OpenMP so that large-scale simulation can be performed by exploiting multicore architecture.

\begin{acks}
This research is partially supported by Intel and NSF Joint Research Center on Computer Assisted Programming for Heterogeneous Architectures (CAPA) (CCF-1723773). We are grateful to Xilinx for the software and the hardware donation. We thank Professor Miryung Kim (UCLA), Chaosheng Shi (Xilinx), and Professor Zhiru Zhang (Cornell Univ.) for the helpful discussions and the suggestions. We also thank Janice Wheeler for proofreading this paper.
\end{acks}

\bibliographystyle{ACM-Reference-Format}
\bibliography{sample-bibliography}

\end{document}